\documentclass[sort&compress,AMA,STIX1COL]{WileyNJD-v2}

\articletype{Article Type}%
\usepackage{subfigure}
\usepackage{amsmath,amssymb,amsfonts}

\begin{document}

\title{High-Order Leader-Follower Tracking Control under Limited Information Availability\protect\thanks{This work was supported in part by the U.S. National Science Foundation under Awards CMMI-1763093 and CMMI-1847651.}}

\author[1]{Chuan Yan}

\author[2]{Tao Yang}

\author[1]{Huazhen Fang*}

\authormark{Yan, Yang and Fang}

\address[1]{\orgdiv{Department of Mechanical Engineering}, \orgname{University of Kansas}, \orgaddress{\state{Kansas}, \country{USA}}}

\address[2]{\orgdiv{State  Key  Laboratory  of  Synthetical  Automation  for Process Industries}, \orgname{ Northeastern University}, \orgaddress{\state{Shenyang}, \country{China}}}


\corres{*Huazhen Fang,   \email{fang@ku.edu}}


\abstract[Abstract]{Limited information availability represents a fundamental challenge for control of multi-agent systems,  since an agent  often lacks   sensing  capabilities to measure certain  states of its own and  can  exchange data only with   its neighbors. The challenge becomes even greater when agents  are governed by  high-order dynamics.  The present work is    motivated to conduct control design for linear and nonlinear high-order leader-follower multi-agent systems    in a context where only the first state of an agent is measured.  To address this open challenge, we develop  novel distributed observers to enable followers to reconstruct  unmeasured or unknown quantities about themselves and the leader    and on such a basis, build  observer-based tracking control approaches. We analyze the convergence properties of the proposed approaches and validate their performance through simulation.}


\maketitle

\section{Introduction}\label{sec:introduction}
Cooperative autonomy based on multi-agent systems (MASs)  is  finding ever-increasing application in different fields. This has driven a surge of research  on  distributed cooperative control for different   tasks, including consensus, leader-follower tracking, synchronization, rendezvous, flocking, optimization, learning, formation control, and coverage control~\cite{sun:2021:TCST,Shamma:Wiley:2008,Wang:Wiley:2017,Ren:Springer:2011,Qin:TIE:2017,
Cortes:SICE:2017,Yang:ARC:2019,Yan:CEP:2018,Mou:AUTO:2015,Murray:DSMC:2007}. Most of the current literature considers agents governed by first- or second-order models.  Although  such low-order models are useful as well as  amenable to control design,
they are often found  inadequate to characterize agents with more complex    higher-order dynamics.  It is also neither trivial nor easy to extend  low-order cooperative control designs to  high-order systems. Recent years  have thus seen a growing amount of work   on high-order MAS control synthesis~\cite{Huang:JAS:2014}.

The study~\cite{ren:2007:DSMC} takes a lead in investigating  high-order MAS consensus,   presenting a distributed consensus control algorithm. This subject has since attracted considerable research efforts, with many studies proposed to deal with various challenges, e.g., directed communication topologies~\cite{Wen:SCL:2013,Abdessamaeud:TAC:2018}, switching topologies~\cite{Wen:SCL:2013,Jiang:IJC:2010}, output feedback design~\cite{Seo:AUTO:2009,Xi:AUTO:2012,Qian:TAC:2019,You:2011:IJRNC,Zhou:AUT:2014,wen:2015:TAC},  bipartite consensus~\cite{Valcher:SCL:2014,Hu:IJRNC:2018},  external disturbances~\cite{Liu:IJRNC:2010,Wang:AUTO:2016},  switching or heterogeneous  dynamics~\cite{Mu:TIE:2017}, quantization~\cite{Li:TCYB:2019}, and constrained energy budget~\cite{Yan:CEP:2018}, to   name   a few. Besides consensus, high-order leader-follower tracking has emerged as another  problem of great interest. A basic form of the problem is introduced  in~\cite{ren:2007:DSMC}, which assumes that the leader agent continuously broadcasts its state information to all the followers. A  consensus-based control algorithm is then developed therein to make the followers track the leader. The study~\cite{Yu:TCAS:2011} considers a more general setting where only a subset of the followers can receive information from the leader. It proposes  a leader-follower tracking control method and proves that followers with small degrees must be informed by the leader to ensure tracking convergence. 
Further, nonlinear dynamics constitutes a stronger challenge for high-order   leader-follower tracking.  The investigations~\cite{zhang:2012:AUT, cui:2012:IET}  propose to adaptively estimate   the nonlinearity inherent in an agent's dynamics using neural networks and then offset  it in the control run.  A few well-known nonlinear control techniques, including   backstepping control and iterative nonlinear control, have also found use in addressing tracking problems of nonlinear multi-agent systems~\cite{li:2014:AUT,jin:2016:SCL,sun:2016:IJRNC}.  
In~\cite{Mondal:IFAC:2016}, a finite-time tracking control approach is developed for a high-order nonlinear MAS with actuator saturation, and the work   in~\cite{Wang:TAC:2017} studies  the problem of finite-time  higher-order tracking with mismatched disturbances.

Despite the importance, the above studies  generally assume that a follower can obtain a large amount of information to make  control decisions. For example, it is required in~\cite{ren:2007:DSMC,Yu:TCAS:2011,zhang:2012:AUT,Mondal:IFAC:2016} that a follower   must know all of its own states, all of the states of its neighbors, and if connected with the leader, all of the leader's states. This requirement can be hardly met  in a real world  where   relevant sensors can be unavailable~\cite{Yu:TCAS:2011}. One can also find similar requirements   in  studies about high-order consensus control. This hence motivates us to explore a more realistic   setting---when only the first state of every agent (leader and followers) is  measured.  It is unsurprising that the low information availability will increase the difficulty for   tracking control.    In addition, the present literature often requires that the leader's dynamics is input-free~\cite{zhang:2012:AUT,Wen:TCYB:2017,cui:2012:IET,jin:2016:SCL,li:2014:AUT,sun:2016:IJRNC}, which comes as another limitation in practice.  
To overcome the challenges, we propose to  exploit the notion of observer-based control design and present two major contributions.  First, we design  an observer-based  tracking control approach for linear high-order MASs. We propose a set of distributed observers to compensate for the limited information,    allowing a follower to comprehensively  estimate the leader's maneuver input and states along with its own states.
These observers are then combined with a nominal   controller to form an observer-embedded tracking controller. We further characterize the convergence of tracking when the proposed controller is applied. As the second contribution,  the study extends to the more challenging case when the agents' dynamics is not only high-order but also nonlinear. Extrapolating the design for the linear case, we develop an observer-based tracking control approach and   analyze its convergence.   Compared to the literature, our work is distinct in two aspects. First, the control design requires very limited information---only every agent's first state. Second, the leader's dynamics is input-driven, and the input is only known by the followers that communicate with the leader. 


 Our work  is further related with two lines of research. 1) Leader-follower tracking  via  output feedback, which generally considers a state-space model   and uses local state or parameter  observers to estimate certain unknown quantities~\cite{Chu:2015:IET,SUN:2016:SCL,Sun:2015:IJC,Shi:IJRNC:2017,Zhang:TAC:2011,Peng:TNNLS:2014}. Differing from them, our study designs distributed observers to help  the followers cooperate to infer the leader's input and states,   removing the restriction  that the leader's maneuver input must be either zero or     known by all the followers. 2) Observer-based first- and second-order tracking control. The literature includes   various kinds of observers designed to allow a  follower to estimate its own velocity~\cite{Zhang:SCL:2014,Xu:Neurocomputing:2015},  the disturbances acting on it~\cite{Wang:TAC:2017}, its velocity relative to the leader~\cite{hu:2015:CNSNS},  the leader's input~\cite{Yan:SCL:2019,Yan:ACC:2018,Yan:IJC:2021}, or the leader's velocity~\cite{hong:2008:distributed,Chen:IJC:2014}.   These   results nonetheless cannot be readily generalized to the  high-order MASs. Finally,  the conference version of this study  in~\cite{Yan:CDC:2019} considers only agents with linear high-order  dynamics. We make substantial expansion in this paper to investigate    MASs with generic-form nonlinear high-order  dynamics.

The rest of the paper is organized as follows. Section~\ref{Notation} introduces the notation  about  graph theory. Section~\ref{High-Order-General} formulates the problem of leader-follower tracking control for a linear high-order MAS, develops an observer-based   tracking control approach, and analyzes its convergence. Section~\ref{High-Order-Nonli} proceeds to study the tracking problem for a nonlinear high-order MAS.   Section~\ref{Simulation} then offers a numerical simulation example to validate the proposed design. Finally, Section~\ref{Conclusion} gathers our conclusions.

\section{Notation}\label{Notation}
 
We use a graph to delineate the   information exchange   topology among the leader and followers. First, consider a network composed of $N$ independent followers, the topology of which is modeled as an undirected graph. The follower graph is expressed as $\mathcal{G}=(\mathcal{V},\mathcal{E})$, where $\mathcal{V}=\{1, 2, \cdots, N\}$ is the node set and $\mathcal{E}\subseteq \mathcal{V}\times \mathcal{V}$ is the edge set  that contains unordered pairs of nodes. A path is a sequence of connected edges in a graph. The follower graph is connected if there is a path between every pair of nodes.
The neighbor set of node $i$ is denoted as $\mathcal{N}_i$. 
The adjacency
matrix of $\mathcal{G}$ is  $A=[a_{ij}]\in\mathbb{R}^{N \times N}$, which has non-negative elements. The element $a_{ij}>0 $  if  and only if $(i,j)\in
\mathcal{E}$, and moreover,
$a_{ii}=0$ for all $i \in \mathcal{V}$. For the Laplacian matrix
$L=[l_{ij}]\in\mathbb{R}^{N \times N}$, $l_{ij}=-a_{ij}$ if $i\neq j$ and $ l_{ii}=\sum_{k\in \mathcal{N}_{i}}a_{ik}$.
The leader is numbered as node $0$ and can send information to its neighboring followers. Then, we define a graph $\bar{\mathcal{G}} = (\bar{\mathcal{V}}, \bar{\mathcal{E}} )$ for the entire network,  where $\bar{\mathcal{V}} = \{0\} \cup  \mathcal{V}$, and $\bar{\mathcal{E}} \subseteq  \bar{\mathcal{V}} \times \bar{\mathcal{V}}$ is the edge set for all nodes.  The leader is globally reachable in $\bar{\mathcal{G}}$ if there is a path in graph $\bar{\mathcal{G}}$ from every node 0 to node $i$.
We denote the leader adjacency matrix associated with $\bar{\mathcal{G}}$ by $B=\mathrm{diag}(b_1,\ldots,b_N)$, where $b_i>0$ if the leader is a neighbor of agent $i$ and $b_i=0$ otherwise.

\section{Leader-Follower Tracking with Linear High-Order Dynamics}\label{High-Order-General}
In this section, we   show the leader-follower tracking problem for   linear high-order MASs and   develop an observer-based tracking control method  as well as   convergence analysis.

\subsection{Problem Formulation}

Consider an MAS composed of $N+1$ agents, among which agent $0$ is   the leader and agents  $1$ to $N$ are followers. Each agent has $l$th-order dynamics   ($l\geq 3$)  expressed as
\begin{subequations}\label{High-Dynamics}
\begin{align}
\dot{x}_{i,m}&=x_{i,m+1},\quad m=1,2,\ldots,l-1,\\
\dot{x}_{i,l}&=u_{i},
\end{align}
\end{subequations}
for $i=0,1,\ldots,N$,
where $x_{i,m}\in \mathbb{R}$ is the $m$th state of agent $i$, and $u_i$   the maneuver  input. The objective   is to design a distributed control law  $u_i$  such that   follower $i$  for $i=1,2,\ldots,N$ can convergently track the leader with $\lim_{t\to \infty}|x_{i,m}(t)-x_{0,m}(t)|=0$ for $m=1,2,\ldots,l$.

Here, we assume that only $x_{i,1}$  for   $i=0,1,\ldots,N$ is available. That is, only the first state of an agent  is measured, regardless of whether it is the leader or a follower. This assumption considerably relaxes the usual requirement in the literature that substantial states of an agent must be measured. 
However, it also implies that   the accessible information about the agents is rather limited, which makes it more challenging to  design  an effective distributed tracking controller.

\subsection{The Proposed Algorithm}

We   develop an observer-based control algorithm to enable convergent tracking in the above setting.
To begin with, we consider the following controller  for follower $i$:
\begin{align}\label{Controller}
u_i=&-k_{1}\left[\sum_{j\in \mathcal{N}_i}a_{ij}(x_{i,1}-x_{j,1})+b_i(x_{i,1}-x_{0,1})\right]
-\sum_{m=2}^{l}k_{m}(\hat{x}_{i,m}-\hat{x}_{0,i,m})+\hat{u}_{0,i},
\end{align}
for $i=1,\ldots,N$, where $k_m$ for $m=1,2,\ldots,l$ are   gain parameters, $\hat{x}_{0,i,m}$ and $\hat{u}_{0,i}$ are follower $i$'s estimates of the leader's state $x_{0,m}$ and input $u_0$, respectively, and $\hat{x}_{i,m}$ is follower's estimate  of its  own state  $x_{i,m}$. The  motivation underlying~\eqref{Controller} is to drive follower $i$ toward its neighbors and the leader simultaneously. When all the followers do this, they can  track the leader in a collective manner.  Next, we design the   observers so as to obtain the   estimates as needed in~\eqref{Controller}.

A distributed input observer is first introduced to  enable the followers to estimate  $u_0$, which is given by
\begin{subequations}\label{U0-Estimate}
\begin{align}
\label{U0-Estimate-1}
&\dot{\hat{u}}_{0,i} =-\sum_{j \in \mathcal{N}_{i}}a_{ij}(\hat{u}_{0,i}-\hat{u}_{0,j})-b_{i}
(\hat{u}_{0,i}-u_{0})
-d_i\cdot\mathrm{sgn}\left[\sum_{j \in \mathcal{N}_{i}}a_{ij}
(\hat{u}_{0,i}-\hat{u}_{0,j})+b_{i}
(\hat{u}_{0,i}-u_{0})\right],\\   \label{U0-Estimate-2}
&\dot{d}_i  =\tau_{i}\left|\sum_{j \in   \mathcal{N}_{i}}a_{ij}(\hat{u}_{0,i}-\hat{u}_{0,j})+b_{i}(\hat{u}_{0,i}-u_{0})\right|,
\end{align}
\end{subequations}
for $i=1,\ldots,N$, where $d_i$ is an adaptive gain, and $\tau_i$  a positive scalar. 
  For~\eqref{U0-Estimate-1}, the first two terms on its right-hand side are used to drive the input estimation $\hat{u}_{0,i}$ to approach   $u_0$ while maintaining consistency with the neighboring followers; further, its  third term  serves as further correction with an adaptive gain   given in~\eqref{U0-Estimate-2} to enhance the convergence.

We further propose another distributed observer for the followers to   estimate $x_{0,m}$ for $m=2,3,\ldots,l$:
\begin{subequations}\label{X0m-Estimate}
\begin{align}
\dot{z}_{0,i,2}&=-b_{i}c_{0,2}z_{0,i,2}-b_{i}^{2}c_{0,2}^2x_{0,1} 
-c_{0,2}\sum_{j \in \mathcal{N}_{i}}a_{ij}(\hat{x}_{0,i,2}-\hat{x}_{0,j,2})+\hat{x}_{0,i,3}, \label{X0m-Estimate-a}\\
\hat{x}_{0,i,2}&=z_{0,i,2}+b_{i}c_{0,2}x_{0,1},\label{X0m-Estimate-b}\\
\dot{z}_{0,i,m}&=-c_{0,m}z_{0,i,m}-c_{0,m}^2\hat{x}_{0,i,m-1}+\hat{x}_{0,i,m+1},\label{X0m-Estimate-c}\\
\hat{x}_{0,i,m}&=z_{0,i,m}+c_{0,m}\hat{x}_{0,i,m-1}, \ \ m=3,4,\ldots,l-1,\\
\dot{z}_{0,i,l}&=-c_{0,l}z_{0,i,l}-c_{0,l}^2\hat{x}_{0,i,l-1}+\hat{u}_{0,i},\\
\hat{x}_{0,i,l}&=z_{0,i,l}+c_{0,l}\hat{x}_{0,i,l-1}, \label{X0m-Estimate-f}
\end{align}
\end{subequations}
for $i=1,\ldots,N$, where $z_{0,i,m}$ and $c_{0,m}$ for $m=2,3,\ldots,l$ are the observer's  internal states  and   gain parameters, respectively.
The development of~\eqref{X0m-Estimate} is inspired by~\cite{yang:2013:ITIE}, in which a centralized disturbance observer
is designed for a single plant. Significantly  transforming the original design, we develop the above observer, which has  a  distributed  structure that is uniquely suitable for the considered MAS setting.  Here, the observer in~\eqref{X0m-Estimate-a}-\eqref{X0m-Estimate-b} attempts  to estimate the leader's second state $x_{0,2}$. In~\eqref{X0m-Estimate-a}, $-\sum_{j \in \mathcal{N}_{i}}a_{ij}(\hat{x}_{0,i,2}-\hat{x}_{0,j,2})$ tries to keep  the estimation consistent between follower $i$ and its neighbors; then,~\eqref{X0m-Estimate-b} performs the estimation to obtain $\hat{x}_{0,i,2}$ using the internal variable $z_{0,i,2}$ and the  leader's first state $x_{0,1}$. Following similar lines in~\eqref{X0m-Estimate-a}-\eqref{X0m-Estimate-b}, the observers in~\eqref{X0m-Estimate-c}-\eqref{X0m-Estimate-f} are designed to estimate the leader's rest higher-order states.


Finally, we design the following observer such that follower $i$ can estimate its own states $x_{i,m}$ for $m=2,3,\ldots,l$:
\begin{subequations}\label{Xim-Estimate}
\begin{align}
\dot{z}_{i,2}&=-r_2z_{i,2}-r_2^2x_{i,1}+\hat{x}_{i,3}, \\
\hat{x}_{i,2}&=z_{i,2}+r_2x_{i,1},\\
\dot{z}_{i,m}&=-r_mz_{i,m}-r_m^2\hat{x}_{i,m-1}+\hat{x}_{i,m+1},\\
\hat{x}_{i,m}&=z_{i,m}+r_m\hat{x}_{i,m-1}, \ m=3,4,\ldots,l-1,\\
\dot{z}_{i,l}&=-r_lz_{i,l}-r_l^2\hat{x}_{i,l-1}+u_{i},\\
\hat{x}_{i,l}&=z_{i,l}+r_l\hat{x}_{i,l-1},
\end{align}
\end{subequations}
for $i=1,\ldots,N$, where $z_{i,m}$ and $r_{i,m}$ for $m=2,3,\ldots,l$ are the  internal states and   gain parameters, respectively.

Putting together~\eqref{Controller}-\eqref{Xim-Estimate}, we   obtain a distributed observer-based control algorithm to achieve  high-order leader-follower tracking. Its convergence is analyzed in Section~\ref{Convergence}.

\begin{remark}
We   highlight a comparison between the proposed approach and  the study of  output-feedback leader-following tracking control in~\cite{Sun:2015:IJC, Chu:2015:IET,SUN:2016:SCL,Shi:IJRNC:2017,Zhang:TAC:2011,Peng:TNNLS:2014}. These references use different observers   to help a follower  estimate its  own states or certain parameters. These observers  are local observers as they are designed  to estimate local unknown quantities. Compared with them, the proposed approach focuses more on distributed observer design---the observers in~\eqref{U0-Estimate}-\eqref{X0m-Estimate} have a distributed structure to enable the followers to   collectively infer the leader's input and states by information exchange. This new design   allows the followers to keep tracking the input-driven leader, setting it apart from the references that restrictively require the leader to be input-free or the followers to have at least certain knowledge of the leader's input.
\hfill$\bullet$
\end{remark}

\subsection{Convergence Analysis}\label{Convergence}
This section characterizes the convergence property for the  algorithm proposed above. Before proceeding further, we make the following  assumption:

\begin{assumption}\label{u0-constraint}
The leader's input $u_0\in \mathcal{C}^1$ with $|\dot{u}_0|\leq w < \infty$, where $  w$ is unknown.
\end{assumption}

This   assumption is reasonable and justifiable due to  the fact that   control actuations are usually smooth and subject to ramp-up/down limits.

As the convergence depends on the estimation and tracking   errors, we   lay out the definitions of these errors first. 
For the observer in~\eqref{U0-Estimate},  we define the estimation error as $e_{u,i}=\hat{u}_{0,i}-u_0$. Its dynamics is
\begin{align}\label{U0-Close-Loop}
\dot{e}_{u,i}
&=-b_ie_{u,i}-\sum_{j \in \mathcal{N}_{i}}a_{ij}(e_{u,i}-e_{u,j})-d_i\cdot\mathrm{sgn}\left[\sum_{j \in \mathcal{N}_{i}}a_{ij}(e_{u,i}-e_{u,j})+b_ie_{u,i}\right]-\dot{u}_0.
\end{align}
Further, define $e_u=\left[\begin{matrix} e_{u,1}& e_{u,2} &\cdots & e_{u,N}\end{matrix}\right]^{\top}$. It then follows from~\eqref{U0-Close-Loop} that
\begin{align}\label{eu-dynamics}
\dot{e}_{u}=-He_u-D\cdot\mathrm{sgn}(He_u)-\dot{u}_0\mathbf{1},
\end{align}
where $H=B+L$ and $D=\mathrm{diag}\{d_{1},\ldots, d_{N}\}$.
It is seen that the signum function term in~\eqref{eu-dynamics} is discontinuous, measurable and locally bounded. Therefore,~\eqref{eu-dynamics} admits a Filippov solution according to~\cite{clarke:1983:NY}, which is governed by the differential inclusion $\mathcal{K}[\cdot]$:
\begin{align}\label{eu-dynam}
\dot{e}_{u}\in^{a.e.} \mathcal{K}\left[-He_u-D\cdot\mathrm{sgn}(He_u)-\dot{u}_0\mathbf{1}\right].
\end{align}

Now, we consider the  observer in~\eqref{X0m-Estimate}. Defining   $e_{0x,i,m}=\hat{x}_{0,i,m}-x_{0,m}$, we have
\begin{subequations}\label{e0xi-Dynamics}
\begin{align} 
\dot{e}_{0x,i,2}&=-c_{0,2}b_ie_{0x,i,2}+e_{0x,i,3}c_{0,2}\sum_{j \in  \mathcal{N}_{i}}a_{ij}({e}_{0x,i,2}-{e}_{0x,j,2}),\\
\dot{e}_{0x,i,m} 
&=-c_{0,m}c_{0,2}b_ie_{0x,i,2}+e_{0x,i,m+1} -c_{0,m}c_{0,2}\sum_{j \in \mathcal{N}_{i}}a_{ij}({e}_{0x,i,2}-{e}_{0x,j,2}), \quad m=3,4,\ldots,l-1,\\
\dot{e}_{0x,i,l} 
&=-c_{0,l}c_{0,2}b_ie_{0x,i,2}+e_{u,i}-c_{0,l}c_{0,2}\sum_{j \in
\mathcal{N}_{i}}a_{ij}(e_{0x,i,2}-e_{0x,j,2}).
\end{align}
\end{subequations}
Define $e_{0x,m}=\left[\begin{matrix} e_{0x,1,m}& e_{0x,2,m} &\cdots & e_{0x,N,m}\end{matrix}\right]^{\top}$ and $e_{0x}=\left[\begin{matrix} e_{0x,2}^\top& e_{0x,3}^\top &\cdots & e_{0x,l}^\top\end{matrix}\right]^{\top}$. Then,~\eqref{e0xi-Dynamics} can be written into a compact form as below:
\begin{align}\label{e0x-Dynamics}
\dot{e}_{0x}=F_1e_{0x}+\ell_1,
\end{align}
where
\begin{align*}
F_1&=\left[\begin{matrix}-c_{0,2}H & I& 0& \cdots& 0\\
\vdots& 0& \ddots& \ddots& \vdots\\
\vdots& \vdots& \ddots& \ddots& 0\\
-c_{0,l-1}c_{0,2}H& 0&\cdots &0 & I\\
-c_{0,l}c_{0,2}H& 0& \cdots& \cdots & 0
\end{matrix}\right], \
\ell_1=\left[\begin{matrix} 0\\
\vdots\\
0\\
e_u  \end{matrix}\right].
\end{align*}

Proceeding further, we define   $e_{x,i,m}=\hat{x}_{i,m}-x_{i,m}$ for the observer in~\eqref{Xim-Estimate} and have
\begin{align*}
\dot{e}_{x,i,2}
&=-r_{2}e_{x,i,2}+{e}_{x,i,3},\\
\dot{e}_{x,i,m}
&= -r_{m}r_{2}e_{x,i,2} +{e}_{x,i,m+1}, \ \ m=3,4,\ldots,l-1,\\
\dot{e}_{x,i,l}
&= -r_{l}r_{2}e_{x,i,2}.
\end{align*}
Define $e_{x,m}=\left[\begin{matrix} e_{x,1,m}& e_{x,2,m} &\cdots & e_{x,N,m}\end{matrix}\right]^{\top}$ for $m=2,3,\ldots,l$ and $e_{x}=\left[\begin{matrix} e_{x,2}^\top& e_{x,3}^\top &\cdots & e_{x,l}^\top\end{matrix}\right]^{\top}$. Then,
\begin{align}\label{ex-Dynamics}
\dot{e}_{x}=F_2e_{x},
\end{align}
where
\begin{align*}
F_2=\left[\begin{matrix}-r_{2}I & I& 0& \cdots& 0\\
\vdots& 0& \ddots& \ddots& \vdots\\
\vdots& \vdots& \ddots& \ddots& 0\\
-r_{l-1}r_{2}I& 0&\cdots &0 & I\\
-r_{l}r_{2}I& 0& \cdots& \cdots & 0
\end{matrix}\right].
\end{align*}

Finally,  to investigate  the state tracking error, we define it as $e_{i,m}=x_{i,m}-x_{0,m}$. The dynamics of $e_{i,m}$ is
\begin{align*}
\dot e_{i,m} &=e_{i,m+1},  \ \ m=1,2,\ldots,l-1 \\ \nonumber
\dot{e}_{i,l} &=-k_{1}\left[\sum_{j\in \mathcal{N}_i}a_{ij}(e_{i,1}-e_{j,1})+b_i e_{i,1} \right]-\sum_{m=2}^{l}k_m e_{i,m} -\sum_{m=2}^{l}k_{m}(e_{x,i,m} -e_{0x,i,m})+e_{u,i}.
\end{align*}
Define
$e_{m}=\left[\begin{matrix} e_{1,m}& e_{2,m} &\cdots & e_{N,m}\end{matrix}\right]^{\top}$ for $m=1,2,\ldots,l$, and $e=\left[\begin{matrix} e_{1}^\top& e_{2}^\top &\cdots & e_{l}^\top\end{matrix}\right]^{\top}$. Here, $e$ is the global tracking error with its dynamics governed by
\begin{align}\label{e-Dynamics}
\dot{e}=F_3e+\ell_3,
\end{align}
where
\begin{align*}
F_3&=\left[\begin{matrix}0 & I& 0& \cdots& 0\\
\vdots& 0& \ddots& \ddots& \vdots\\
\vdots& \vdots& \ddots& \ddots& 0\\
0& 0&\cdots &0 & I\\
-k_{1}H& -k_2I& \cdots& \cdots & -k_lI
\end{matrix}\right],
\ell_3 =\left[\begin{matrix}0\\ \vdots\\
0\\ -\sum_{m=2}^{l}k_m(e_{x,m}-e_{0x,m})+e_u\end{matrix}\right].
\end{align*}

As a main result, the following theorem outlines the convergence property of the proposed algorithm.

\begin{theorem}
Suppose that  the controller in~\eqref{Controller}-\eqref{Xim-Estimate} is applied to~\eqref{High-Dynamics}.  If  Assumption~\ref{u0-constraint} holds, then
\begin{align}\label{u0-analysis}
\lim_{t\to\infty} e_u(t) = 0.
\end{align}
Further, $\lim\limits_{t\to\infty} e_{0x}(t) = 0$ if there exist $c_{0,2}, c_{0,3}, \ldots, c_{0,l}>0$ such that the   polynomials
\begin{align}\label{Poly-F1}
h_i(s) = s^{l-1}+c_{0,2}s^{l-2}\lambda_i(H)
+c_{0,2}\lambda_i(H) \sum_{z=0}^{l-3} c_{0,l-z} s^{z} 
\end{align}
for $i=1,2,\ldots,N$ are Hurwitz stable, and
$\lim_{t\to\infty} e_x(t) = 0$ if there exist  $r_2,r_3,\ldots, r_l>0$ such that the polynomial 
\begin{align}\label{Poly-F2}
s^{l-1}+r_{2}s^{l-2}
+ r_{2} \sum_{z=0}^{l-3} r_{l-z} s^{z}
\end{align}
is Hurwitz stable. Finally, the global tracking error $\lim_{t\to \infty}e(t)=0$ if there exist   $k_m$ for $m=1,2,\ldots,l$ such that the polynomials
\begin{align}\label{F3-Stable}
s^{l}+k_1\lambda_i(H)
+\sum_{z=2}^{l}s^{z-1}k_z
\end{align}
for $i=1,2,\ldots,N$  are Hurwitz stable.
\end{theorem} 

{\noindent\bf Proof:} To prove~\eqref{u0-analysis}, 
let us consider a Lyapunov functional candidate, $V_1=\bar{V}_1(e_u)+\tilde{V}_1 (d_i)$, where
\begin{align*}
\bar{V}_1(e_u)=\frac{1}{2}e_u^\top He_u,\
\tilde{V}_1 (d_i)=\sum_{i=1}^N \frac{(d_i-\beta)^{2}}{2\tau_i},
\end{align*}
with  $\beta\geq w$.
The set-valued Lie derivative of $\bar{V}_1(e_u)$ denoted by $\mathcal{L}\dot{\bar{V}}_1$ along~\eqref{eu-dynam} is given by
\begin{align*}
&\mathcal{L}\dot{\bar{V}}_1=\mathcal{K}\left[-e_u^\top H^2e_u-e_u^\top H D\cdot \mathrm{sgn}(He_u)-e_u^\top H\dot{u}_0\mathbf{1}\right]\\ \nonumber
&= \mathcal{K}\Big[-\sum_{i=1}^N d_i\left(\sum_{j \in \mathcal{N}_{i}}a_{ij}(\hat{u}_{0,i}-\hat{u}_{0,j})+b_{i}(\hat{u}_{0,i}-u_{0})\right)^\top\\ \nonumber
&\quad \cdot\mathrm{sgn}\left(\sum_{j \in \mathcal{N}_{i}}a_{ij}(\hat{u}_{0,i}-\hat{u}_{0,j})+b_{i}(\hat{u}_{0,i}-u_{0})\right) -e_u^\top H^2e_u-e_u^\top H\dot{u}_0\mathbf{1}\Big]\\ \nonumber
&\leq-\sum_{i=1}^N d_i\left|\sum_{j \in \mathcal{N}_{i}}a_{ij}(\hat{u}_{0,i}-\hat{u}_{0,j})+b_{i}(\hat{u}_{0,i}-u_{0})\right| -e_u^\top H^2e_u+w\|He_u\|_1,
\end{align*}
where the fact that $\mathcal{K}[f]=\{f\}$ if $f$ is continuous is used. It then is obtained that $\dot{\bar{V}}_1\in \mathcal{L}\dot{\bar{V}}_1$~\cite{clarke:1983:NY}. Hence,
\begin{align*}
\dot{V}_1&=\dot{\bar{V}}_1+\dot{\tilde{V}}_1=\dot{\bar{V}}_1+\sum_{i=1}^N \frac{(d_i-\beta)\dot{d_i}}{\tau_i}\\ \nonumber
&\leq-\sum_{i=1}^N d_i\left|\sum_{j \in \nonumber \mathcal{N}_{i}}a_{ij}(\hat{u}_{0,i}-\hat{u}_{0,j})+b_{i}(\hat{u}_{0,i}-u_{0})\right|\\ \nonumber
&\quad +\sum_{i=1}^N (d_i-\beta)\left|\sum_{j \in \mathcal{N}_{i}}a_{ij}(\hat{u}_{0,i}-\hat{u}_{0,j})+b_{i}(\hat{u}_{0,i}-u_{0})\right| -e_u^\top H^2e_u+w\|He_u\|_1 \\
&=-e_u^\top H^2e_u-(\beta-w)\|He_u\|_1.
\end{align*}
Note that $H$ is positive definite~\cite{hu:2007:PASMA}.  This, together with $\beta \geq w$, indicates $\dot{V}_1\leq 0$. Thus, $V_1(e_u)$ is non-increasing, implying that $e_u$ and $d_i$ are bounded.
It follows from~\eqref{U0-Estimate-2} that $d_i$ is monotonically increasing. This means that  $d_i$ should  converge to some finite value. In the meantime, $V_1(e_u)$ reaches a finite limit as it is decreasing and lower-bounded by zero. If denoting $s(t) =\int_0^t e_u^\top (\tau)H^2e_u(\tau) d \tau$, we see that $s(t) \leq V_1(0)-V_1(t) $ by integrating $\dot{V}_1(e_u)\leq -e_u^\top H^2e_u$. Hence, $\lim_{t\to\infty} s(t)$ exists and is finite. Due to the boundedness of $e_u$ and $\dot e_u$, $\ddot s$ is also bounded. This implies that $\dot s$ is uniformly continuous. Then, $\lim_{t\to\infty} \dot s(t) = 0$
by Barbalat's Lemma~\cite{khalil:1996:PHNJ}. Therefore, we can obtain $\lim_{t \to \infty}e_u(t)=0$.  

To prove the asymptotic stability of $e_{0x}$, we use the Schur complement, we can find out that the characteristic polynomial of $F_1$ is $  \prod_{i=1}^{N}h_i(s)$, where $h_i(s)$ is shown in~\eqref{Poly-F1}. Note that  $\lim_{t\to\infty} \ell_1(t) = 0$ due to~\eqref{u0-analysis}. Hence, we have  $\lim_{t\to\infty} e_{0x}(t) = 0$ based on the   input-to-state stability (ISS) theory~\cite{khalil:1996:PHNJ}.
Following similar  lines, we can obtain that $\lim_{t\to\infty} e_x(t) = 0$ given~\eqref{Poly-F2}, and further prove that  $\lim_{t\to \infty}e(t)=0$ if~\eqref{F3-Stable} holds. \hfill$\bullet$

\begin{remark}
The proposed controller only requires the neighboring followers to interchange $x_{i,1}$, $\hat u_{0,i}$ and  $\hat x_{0,i,2}$, because the   observers can locally estimate other   quantities necessary for control. This greatly reduces the amount of data to be exchanged between agents and makes the design more advantageous in terms of communication costs. 
\hfill$\bullet$
\end{remark}

\section{High-Order Tracking for Nonlinear Dynamics}\label{High-Order-Nonli}

We have investigated  leader-follower tracking for a linear high-order MAS in the previous section. Given the importance of nonlinear leader-follower MASs, 
this section moves forward to study the case when an MAS has nonlinear high-order dynamics. We will   develop an observer-based tracking control algorithm and analyze its convergence properties.

Suppose that agent $i$'s dynamics is governed by
\begin{subequations}\label{High-Dynamics-Nonli}
\begin{align}
\dot{x}_{i,m}&=x_{i,m+1}+f_m(\overline{x_{i,m}}), \quad m=1,2,\ldots,l-1,\\
\dot{x}_{i,l}&=u_{i}+f_l(\overline{x_{i,l}}),
\end{align}
\end{subequations}
for $i = 0,1,\ldots,N$,
where  $f_m(\overline{x_{i,m}}):\mathbb{R}^m\rightarrow \mathbb{R}$ for $m=1,2,\ldots,l$ are  nonlinear functions with $\overline{x_{i,m}}=(x_{i,1},x_{i,2},\ldots, x_{i,m})$. Following Section~\ref{High-Dynamics}, we assume that only $x_{i,1}$  is measured and continue to hold  Assumption~\ref{u0-constraint}. The control design objective here  is still to enable   convergent tracking, i.e., $\lim_{t\to \infty}|x_{i,m}(t)-x_{0,m}(t)|=0$ for $m=1,2,\ldots,l$ and $i=1,2,\ldots,N$.

To achieve the above objective, we propose the following distributed controller:
\begin{align}\label{Controller-Nonli}
u_i=-k_{1}(x_{i,1}-\hat{x}_{0,i,1})
-\sum_{m=2}^{l}k_{m}(\hat{x}_{i,m}-\hat{x}_{0,i,m})+\hat{u}_{0,i}.
\end{align}
This controller   must be supplemented by corresponding observers. It is noted first that the observer in~\eqref{U0-Estimate} can also be applied  to obtain $\hat u_{0,i}$ here, so we continue to use it   for the distributed estimation of $u_0$. We then construct the following state observer to allow follower $i$ to estimate $x_{0,m}$ for $m=1,2,\ldots,l$:
\begin{subequations}\label{X0m-Estimate-Nonli}
\begin{align}
\dot{\hat{x}}_{0,i,1}&=-c_{0,1}\left[\sum_{j \in \mathcal{N}_{i}}a_{ij}(\hat{x}_{0,i,1}-\hat{x}_{0,j,1})
+b_{i}(\hat{x}_{0,i,1}-x_{0,1})\right] +\hat{x}_{0,i,2}+f_1(\overline{\hat{x}_{0,i,1}}),\\ \nonumber
\dot{z}_{0,i,2}&=-b_{i}c_{0,2}z_{0,i,2}-b_{i}^{2}c_{0,2}^2x_{0,1}+\hat{x}_{0,i,3}
+f_2(\overline{\hat{x}_{0,i,2}})\\
&\quad-c_{0,2}\sum_{j \in \mathcal{N}_{i}}a_{ij}(\hat{x}_{0,i,2}-\hat{x}_{0,j,2})-b_ic_{0,2}f_1(x_{0,1}), \\
\hat{x}_{0,i,2}&=z_{0,i,2}+b_{i}c_{0,2}x_{0,1},\\ 
\dot{z}_{0,i,m}&=-c_{0,m}z_{0,i,m}-c_{0,m}^2\hat{x}_{0,i,m-1}+\hat{x}_{0,i,m+1} +f_m(\overline{\hat{x}_{0,i,m}})-c_{0,m}f_{m-1}(\overline{\hat{x}_{0,i,m-1}}),\\
\hat{x}_{0,i,m}&=z_{0,i,m}+c_{0,m}\hat{x}_{0,i,m-1},\quad  m = 3,4,\ldots,l-1,
\\ 
\dot{z}_{0,i,l}&=-c_{0,l}z_{0,i,l}-c_{0,l}^2\hat{x}_{0,i,l-1}+\hat{u}_{0,i} +f_l(\overline{\hat{x}_{0,i,l}})-c_{0,l}f_{l-1}(\overline{\hat{x}_{0,i,l-1}}),\\
 \hat{x}_{0,i,l}&=z_{0,i,l}+c_{0,l}\hat{x}_{0,i,l-1},
\end{align}
\end{subequations}
where $\overline{\hat{x}_{0,i,m}}=(\hat{x}_{0,i,1},\hat{x}_{0,i,2},\ldots,\hat{x}_{0,i,m})$.  To make a follower  able to estimate its own states, we develop an observer as follows:
\begin{subequations}\label{Xim-Estimate-Nonli}
\begin{align}
\dot{z}_{i,2}&=-r_2z_{i,2}-r_2^2x_{i,1}+\hat{x}_{i,3}+f_2(x_{i,1},\hat{x}_{i,2})-r_2f_1(x_{i,1}), \\
\hat{x}_{i,2}&=z_{i,2}+r_2x_{i,1},\\ 
\dot{z}_{i,m}&=-r_mz_{i,m}-r_m^2\hat{x}_{i,m-1}+\hat{x}_{i,m+1}+f_m(x_{i,1},\overline{\hat{x}_{i,m}})-r_mf_{m-1}(x_{i,1},\overline{\hat{x}_{i,m-1}}),\\
\hat{x}_{i,m}&=z_{i,m}+r_m\hat{x}_{i,m-1},\quad m=3,4,\ldots,l-1,
\\ 
\dot{z}_{i,l}&=-r_lz_{i,l}-r_l^2\hat{x}_{i,l-1}+u_{i} +f_l(x_{i,1},\overline{\hat{x}_{i,l}})-r_lf_{l-1}(x_{i,1},\overline{\hat{x}_{i,l-1}}),\\
\hat{x}_{i,l}&=z_{i,l}+r_l\hat{x}_{i,l-1},
\end{align}
\end{subequations}
where $\overline{\hat{x}_{i,m}}=(\hat{x}_{i,2},\hat{x}_{i,3},\ldots,\hat{x}_{i,m})$.

Integrating the above observers in~\eqref{U0-Estimate},~\eqref{X0m-Estimate-Nonli}-\eqref{Xim-Estimate-Nonli} into  the controller  in~\eqref{Controller-Nonli} will yield  a complete observer-based tracking controller. Before going further to analyze its effectiveness, we make the following  assumption:

\begin{assumption}\label{Lipschitz}
There  exist $\rho_m\geq 0$  such that
\begin{align*}
|f_m( \xi)-f_m(\epsilon )| \leq  \rho_m\|\xi-\epsilon\|, \ m=1,2,\ldots,l,
\end{align*}
 where $\xi, \epsilon \in \mathbb{R}^m$.
\end{assumption}

Assumption~\ref{Lipschitz} implies that the   nonlinear functions  must be of Lipschitz class. It  is commonly used in the literature on   nonlinear MAS control and can be satisfied by many practical systems.

The following theorem shows the main result about convergence    of the proposed controller.

\begin{theorem}\label{Nonli-Stable}
Assume that Assumptions~\ref{u0-constraint} and~\ref{Lipschitz} hold and that the   controller proposed above  is applied to~\eqref{High-Dynamics-Nonli}. The state  tracking error converges to zero, i.e., $\lim_{t\rightarrow \infty}|x_{i,m}(t)-x_{0,m}(t)|=0$ for $m=1,2,\ldots,l$ and $i=1,2,\ldots,N$, if
there exist   $c_{0,m}$, $r_{n}$ and  $k_m$ for $m=1,2,\ldots,l$ and $n=2,\ldots,N$  such that the polynomials~\eqref{Poly-F1},~\eqref{Poly-F2} and
\begin{align}\label{Poly-F6}
\left(s^{l}+\sum_{z=1}^{l}s^{z-1}k_z\right)^N
\end{align}
are Hurwitz stable, and if there exist  matrices $Q_i>0$  and  $\eta_i>0$ for $i=1,2,3$ such that
\begin{subequations}
\begin{align}
&F_4^\top Q_1+Q_1F_4= -\eta_1I,    \label{F4-Sta}   \\
&F_2^\top Q_2+Q_2F_2= -\eta_2I,     \label{F2-Sta}    \\
&F_6^\top Q_3+Q_3F_6= -\eta_3I,    \label{F6-Sta}  \\
&\sum_{i=1}^{l}\|P_{0x,i}\| < \min\left\{\frac{\eta_1}{2\|Q_1\|} ,
\frac{\eta_3}{2\|Q_3\|}\right\},  \label{e0x-Stable}\\
&\sum_{i=2}^{l}\|P_{x,i}\|<\frac{ \eta_2}{2\|Q_2\|},    \label{ex-Stable}
\end{align}
\end{subequations}
where $P_{0x,i}=\mathrm{diag}\{\rho_1I, \rho_2I, \ldots, \rho_iI, 0I, 0I, \ldots, 0I\}$, $P_{x,i}=\mathrm{diag}\{\rho_2I, \rho_3I, \ldots, \rho_iI, 0I, 0I, \ldots, 0I\}$ for $i=1, 2, \ldots, l$, and
\begin{align*}
F_4&=\left[\begin{matrix}-c_{0,1}H&I&0&\cdots&\cdots&0\\
0&-c_{0,2}H & I& 0& \cdots& 0\\
\vdots&\vdots& 0& \ddots& \ddots& \vdots\\
\vdots&\vdots& \vdots& \ddots& \ddots& 0\\
\vdots&-c_{0,l-1}c_{0,2}H& 0&\cdots &0 & I\\
0&-c_{0,l}c_{0,2}H& 0& \cdots& \cdots & 0
\end{matrix}\right], 
F_6=\left[\begin{matrix}0 & I& 0& \cdots& 0\\
\vdots& 0& \ddots& \ddots& \vdots\\
\vdots& \vdots& \ddots& \ddots& 0\\
0& 0&\cdots &0 & I\\
-k_{1}I& -k_2I& \cdots& \cdots & -k_lI
\end{matrix}\right].
\end{align*}

\end{theorem}

{\noindent\bf Proof:}
Let us define  $e_{0x}=\left[\begin{matrix} e_{0x,1}^\top& e_{0x,2}^\top &\cdots & e_{0x,l}^\top\end{matrix}\right]^{\top}$, where  $e_{0x,m}$ follows the same definition as in Section~\ref{High-Order-General}, and define
$
f_{0x,m}=\left[\begin{matrix} \cdots & f_m(\overline{\hat{x}_{0,i,m}})-f_m(\overline{x_{0,m}})& \cdots& \end{matrix} \right]^{\top}$
for $i=1,2,\cdots,N$.
By~\eqref{High-Dynamics-Nonli} and~\eqref{X0m-Estimate-Nonli}, we have
\begin{align}\label{e0x-Dynamics-Nonli}
\dot{e}_{0x}=F_4e_{0x}+\ell_4,
\end{align}
where $\ell_4=\left[\begin{matrix} f_{0x,1}^\top &
\cdots &
f_{0x,l-1}^\top &
f_{0x,l}^\top+e_u^\top  \end{matrix}\right]^\top$.
We  choose a Lyapunov candidate function
\begin{align*}
V_2(e_{0x})=\frac{1}{2}e_{0x}^{\top}Q_1e_{0x},
\end{align*} for which there exist  $\alpha_1, \alpha_2>0$ such that
\begin{align*}
 \alpha_1\|e_{0x}\|^2\leq V_2(e_{0x})\leq \alpha_2\|e_{0x}\|^2.
\end{align*}
By~\eqref{F4-Sta}, we  have
\begin{align*}
\dot{V_2}&=\frac{1}{2}e_{0x}^\top (Q_1F_4+F_4^{\top}Q_1)e_{0x}+e_{0x}^\top Q_1\ell_4 \\\nonumber
&\leq -\frac{1}{2}\eta_1\|e_{0x}\|^{2}+\|e_{0x}\|\|Q_1\|\|\ell_4\|\\
&\leq -\frac{1}{2}\eta_1\|e_{0x}\|^{2}+\|e_{0x}\|\|Q_1\|\big(\|P_{0x,1}e_{0x}\|
+\|P_{0x,2}e_{0x}\|\\
&\quad+
\cdots+\|P_{0x,l}e_{0x}\|+\|e_u\|\big)\\
&=-\left(\frac{1}{2}\eta_1-\|Q_1\|\sum_{i=1}^{l}\|P_{0x,i}\|\right)\|e_{0x}\|^{2}+\|e_{0x}\|\|Q_1\|\|e_u\|,
\end{align*}
Define
\begin{align*}
\sigma_1=\frac{1}{2}\eta_1-\|Q_1\|\sum_{i=1}^{l}\|P_{0x,i}\|,\
\mathcal{X}(\|e_u\|)=\frac{\|Q_1\|\|e_u\|}{\sigma_1\theta_1}
\end{align*}
for any $0<\theta_1<1$. By~\eqref{e0x-Stable}, one can see that $\sigma_1>0$. It can be verified that
\begin{align*}
\|e_{0x}\|\geq \mathcal{X}(\|e_u\|)\Rightarrow \dot{V_2}\leq-\sigma_1(1-\theta_1)\|e_{0x}\|^{2}.
\end{align*}
 Hence, $V_2$ is an ISS-Lyapunov function, implying that  the system~\eqref{e0x-Dynamics-Nonli} is ISS~\cite{khalil:1996:PHNJ}. Then, we have $\lim_{t\to \infty}e_{0x} =0$ since $\lim_{t\to \infty}e_{u} =0$ as indicated in~\eqref{u0-analysis}.

Define
$f_{x,m}=\left[\begin{matrix} \cdots & f_m(x_{i,1},\overline{\hat{x}_{i,m}})-f_m(x_{i,1},\overline{x_{i,m}})&   \cdots \end{matrix}\right]^{\top}$
for $i=1,2,\ldots,N$, 
and continue to adopt $e_{x}$ as defined  in~\eqref{ex-Dynamics}.                       According to~\eqref{Xim-Estimate-Nonli}, its dynamics can be expressed as
\begin{align}\label{ex-Dynamics-Nonli}
\dot{e}_{x}=F_2e_{x}+\ell_5,
\end{align}
where $F_2$ was defined in~\eqref{ex-Dynamics}, and $\ell_5=\left[\begin{matrix} f_{x,2}^\top &
\cdots &
f_{x,l}^\top \end{matrix}\right]^\top$. Following similar lines to the above, we can prove that $\lim_{t\to \infty}e_x = 0$ if~\eqref{F2-Sta} and \eqref{ex-Stable} hold.

We proceed to consider the global tracking error when the controller in~\eqref{Controller-Nonli} is applied. We define $f_{m}=\left[\begin{matrix} \cdots &  f_m(\overline{x_{i,m}})-f_m(\overline{x_{0,m}}) & \cdots\end{matrix}\right]^{\top}$ for $i=1,2,\cdots,N$.
The dynamics of the tracking error $e_{i,m} = {x}_{i,m}-{x}_{0,m}$ is
\begin{subequations}\label{Closed-Loop-i-Nonli}
\begin{align}
\dot{e}_{i,m}& =e_{i,m+1}+f_m(\overline{x_{i,m}})-f_m(\overline{x_{0,m}}),   \\
\dot{e}_{i,l}
&=-\sum_{m=1}^{l}k_m e_{i,m}-\sum_{m=2}^{l}k_me_{x,m}+\sum_{m=1}^{l}k_me_{0x,m}+e_{u,i}+f_l(\overline{x_{i,l}})-f_l(\overline{x_{0,l}}),
\end{align}
\end{subequations}
for $m=1,2,\ldots,l-1$ and $i=1,2,\ldots, N$. The notation of $e$ in~\eqref{e-Dynamics} is still adopted here.
Now, combining~\eqref{X0m-Estimate-Nonli},~\eqref{Xim-Estimate-Nonli} and~\eqref{Closed-Loop-i-Nonli}, the closed-loop system is indicated into a compact structure as below:
\begin{align}\label{e-Dynamics-Nonli}
\dot{e}=F_6e+\ell_6+\ell_7,
\end{align}
where $\ell_6=\left[\begin{matrix} f_1^\top &
\cdots &
f_{l}^\top \end{matrix}\right]^\top$ and
\begin{align*}
\ell_7 =\left[\begin{matrix}0\\ \vdots\\
0\\ -\sum_{m=2}^{l}k_me_{x,m}+\sum_{m=1}^{l}k_me_{0x,m}+e_u\end{matrix}\right].
\end{align*}
For~\eqref{e-Dynamics-Nonli}, we can use the ISS theory to prove that it is asymptotically stable if~\eqref{F6-Sta}-\eqref{e0x-Stable} hold. Therefore, $\lim_{t\to \infty}e(t)=0$ is established as $\lim_{t\to \infty}\ell_7(t)=0$ due to~\eqref{eu-dynamics},~\eqref{e0x-Dynamics-Nonli} and~\eqref{ex-Dynamics-Nonli}. We conclude that $\lim_{t\to \infty}|x_{i,m}(t)-x_{0,m}(t)|=0$.  Finally, we highlight that the Hurwitz stable polynomials~\eqref{Poly-F1},~\eqref{Poly-F2} and~\eqref{Poly-F6} would ensure the existence of solutions for the Lyapunov equations in~\eqref{F4-Sta}-\eqref{F6-Sta}.  This concludes  the proof.
\hfill$\bullet$

\begin{remark}
  Theorem~\ref{Nonli-Stable} can be briefly explained as follows.   The conditions~\eqref{F4-Sta} and~\eqref{e0x-Stable} ensure that the distributed observers in~\eqref{X0m-Estimate-Nonli} are asymptotically stable; the conditions~\eqref{F2-Sta} and~\eqref{ex-Stable} ensure that the local state observers in~\eqref{Xim-Estimate-Nonli} are asymptotically stable; further, the conditions~\eqref{F6-Sta} and~\eqref{e0x-Stable} ensure that the global closed-loop tracking errors converge to zero. 
When the polynomials~\eqref{Poly-F1},~\eqref{Poly-F2} and~\eqref{Poly-F6} are Hurwitz  stable,  $F_4$, $F_2$, and $F_6$ will be stable,  and then the Lyapunov equations in~\eqref{F4-Sta}-\eqref{F6-Sta} will admit solutions. 
\end{remark} 

\begin{remark}
The above design can be extended to the case when the nonlinear functions $f_m (\cdot)$ is unknown but admits approximation by a known function with bounded error. Specifically,  suppose that  there exist $g_m(\cdot)$ and $\varphi_m \geq 0$ such that
\begin{align*}
\left| g_m(\phi) - f_m (\phi) \right| \leq \varphi_m, \ m=1, 2,\ldots,l,
\end{align*}
 for any $\phi \in  \mathbb{R}^m$. We then can replace $f_m(\cdot)$ in~\eqref{X0m-Estimate-Nonli}-\eqref{Xim-Estimate-Nonli} by $g_m(\cdot) $ and  obtain a tracking controller based on the approximate nonlinearity. It can be proven that this controller will lead to bounded-error tracking under certain mild conditions. The analysis is omitted here for the sake of space.
\end{remark}

\begin{figure}[t]
\centering
 \includegraphics[scale=0.4]{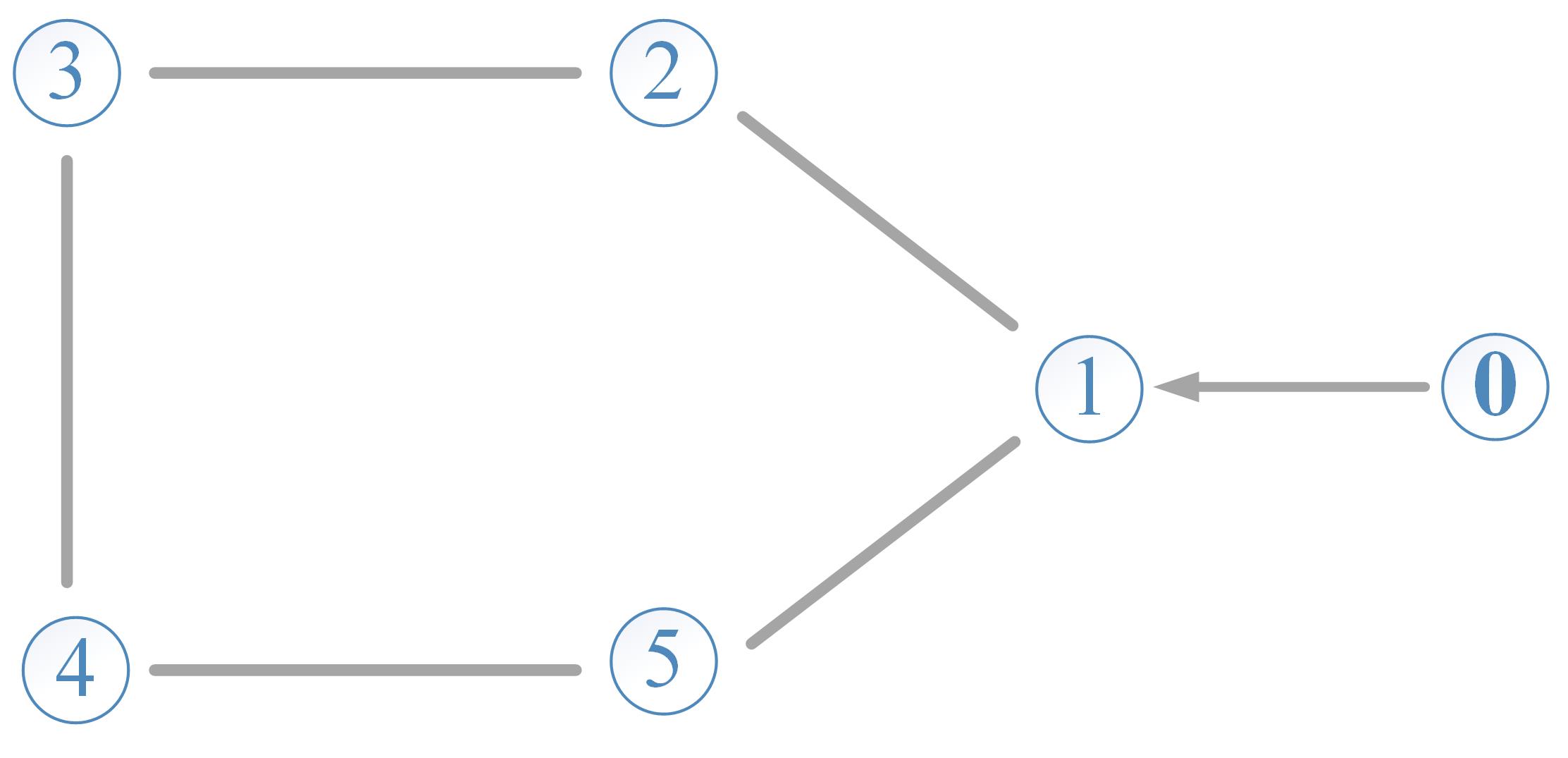}
 \caption{Topology of the MAS in the simulation.}
\label{MAS-topology-sim}
\end{figure}

\begin{figure*}[t]
\centering
\subfigure[]{
\includegraphics[trim={3mm 16mm 11mm
 20mm},clip,width=0.325\linewidth]{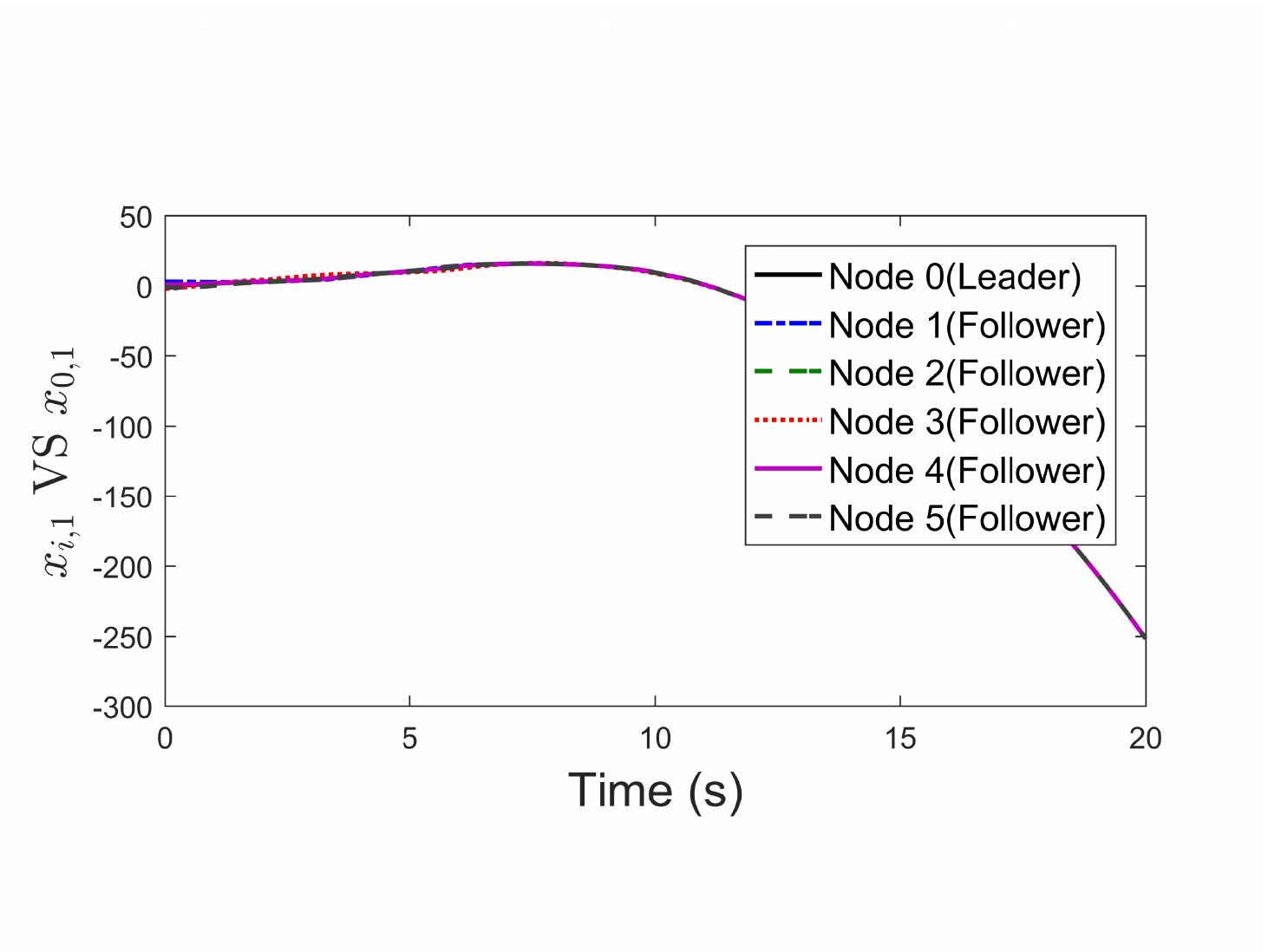}\label{X01-Nonli}} 
\subfigure[]{
\includegraphics[trim={5mm 16mm 12mm
 20mm},clip,width=0.325\linewidth]{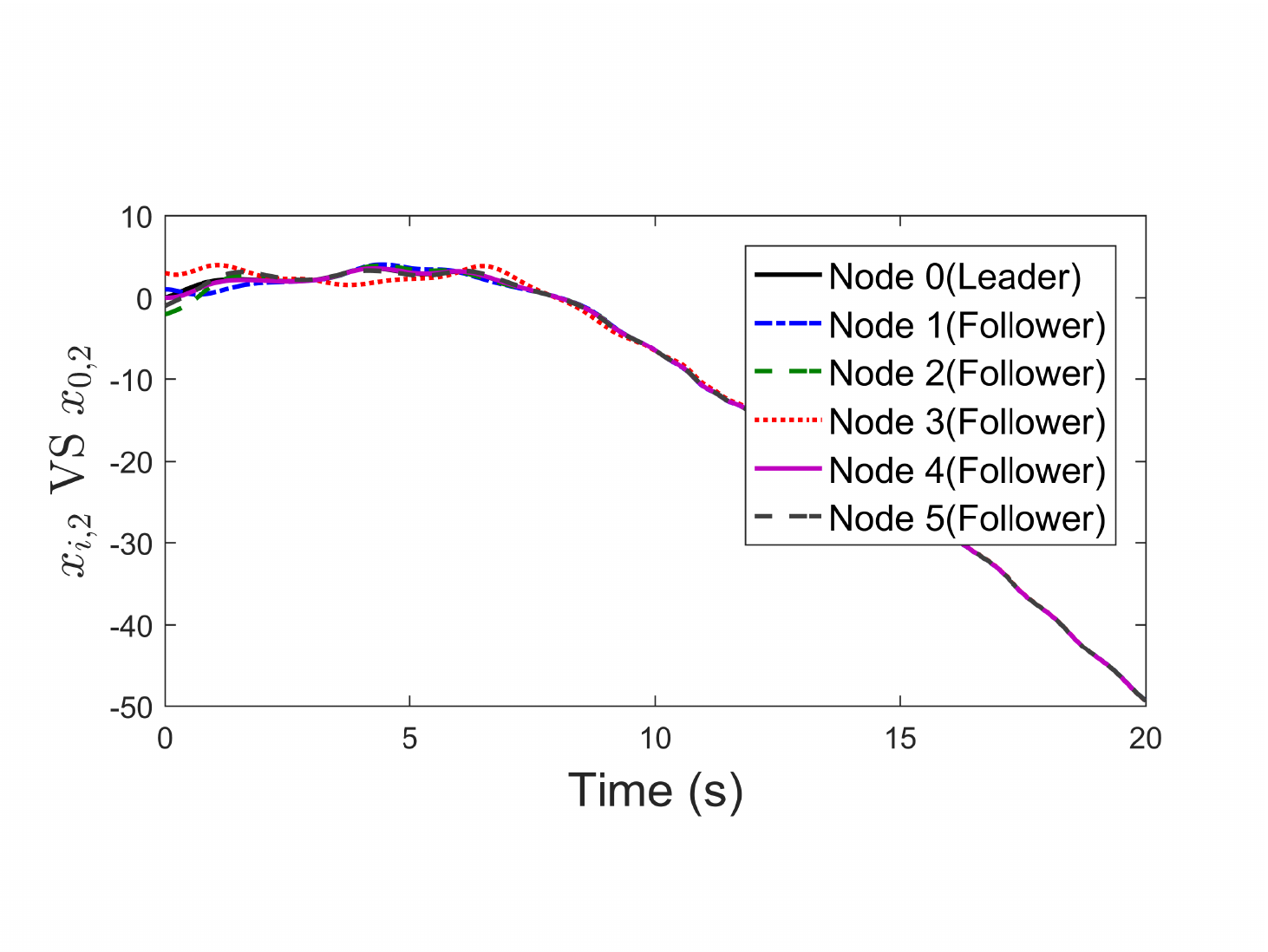}\label{X02-Nonli}}  
 \subfigure[]{
\includegraphics[trim={5mm 16mm 11mm
20mm},clip,width=0.325\linewidth]{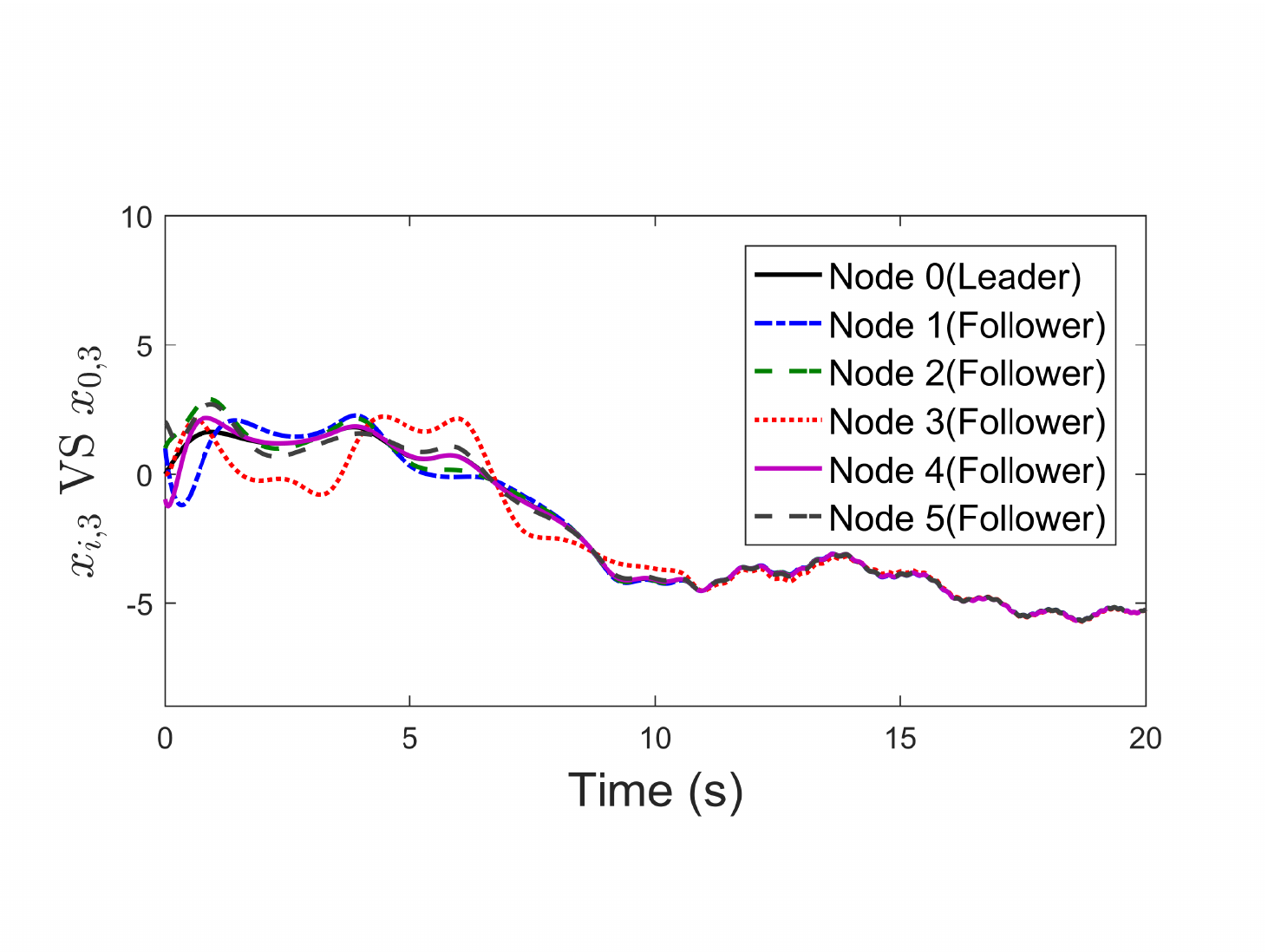}\label{X03-Nonli}}\\
 \subfigure[]{
\includegraphics[trim={4mm 16mm 11mm
 20mm},clip,width=0.325\linewidth]{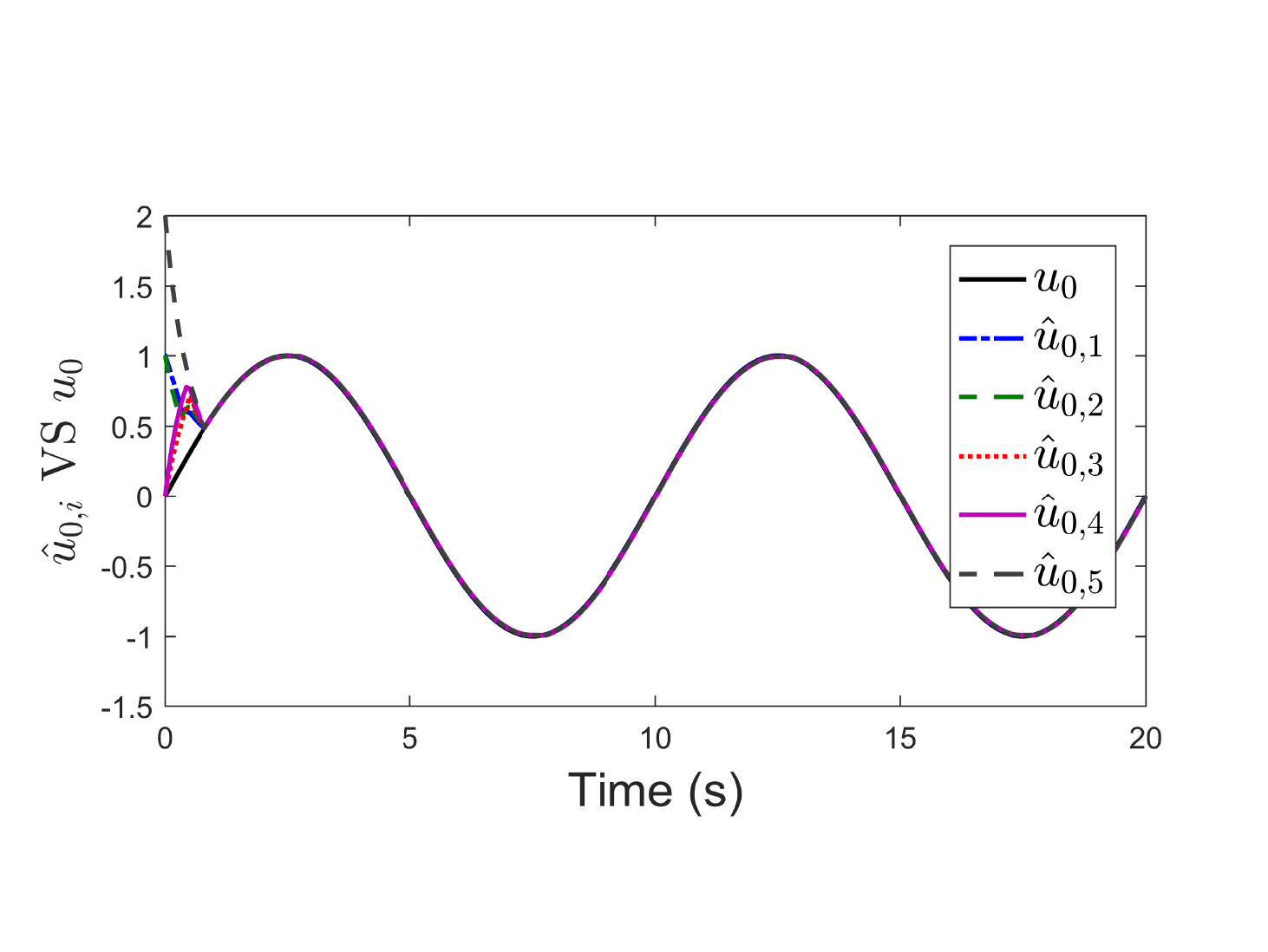}\label{U0i-Nonli}}
 \subfigure[]{
\includegraphics[trim={3mm 16mm 11mm
20mm},clip,width=0.325\linewidth]{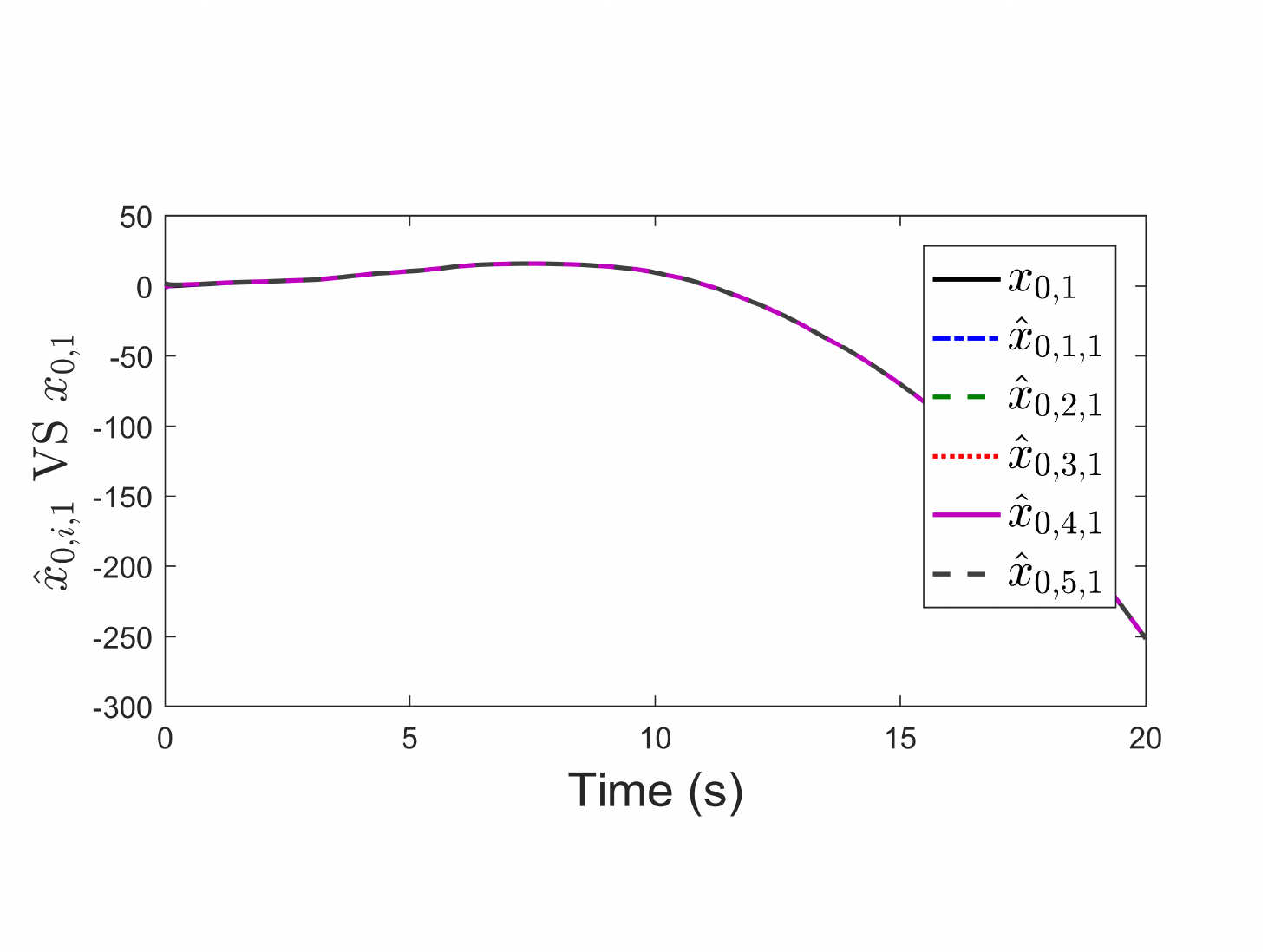}\label{X0i1-Nonli}} 
 \subfigure[]{
\includegraphics[trim={5mm 16mm 11mm
20mm},clip,width=0.325\linewidth]{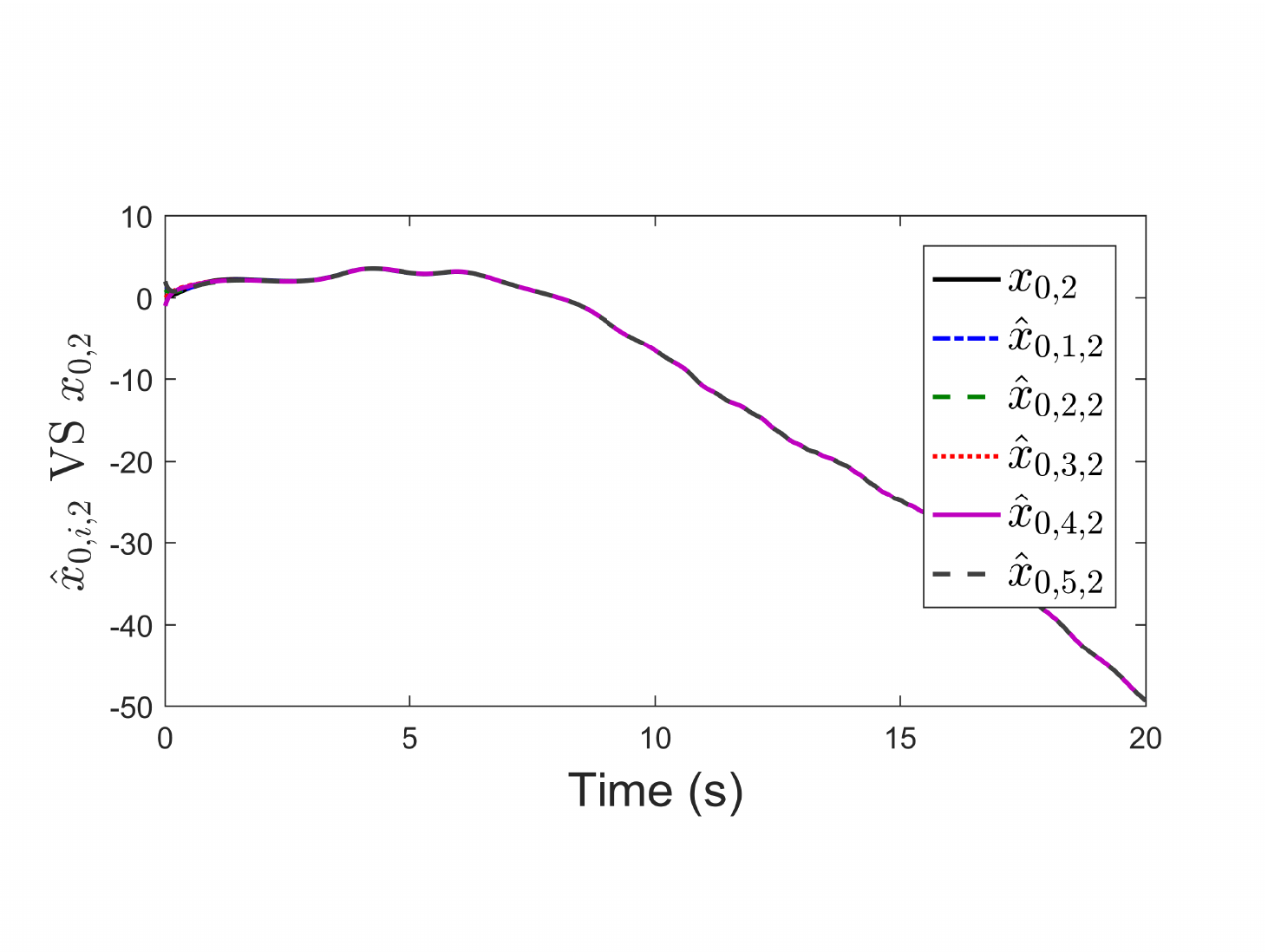}\label{X0i2-Nonli}}\\
\subfigure[]{
\includegraphics[trim={5mm 16mm 12mm
 20mm},clip,width=0.325\linewidth]{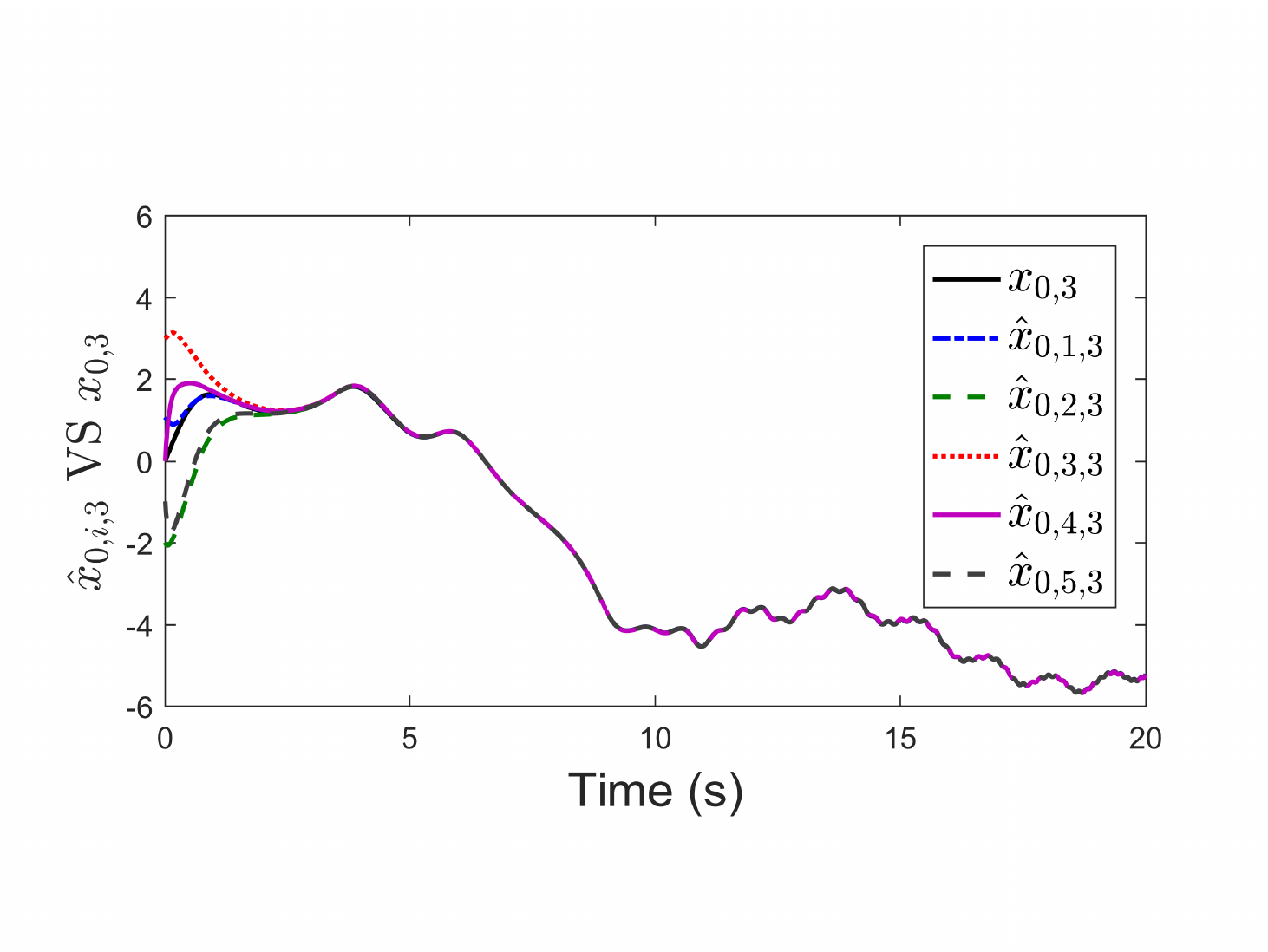}\label{X0i3-Nonli}} 
\subfigure[]{
\includegraphics[trim={5mm 16mm 11mm
 20mm},clip,width=0.325\linewidth]{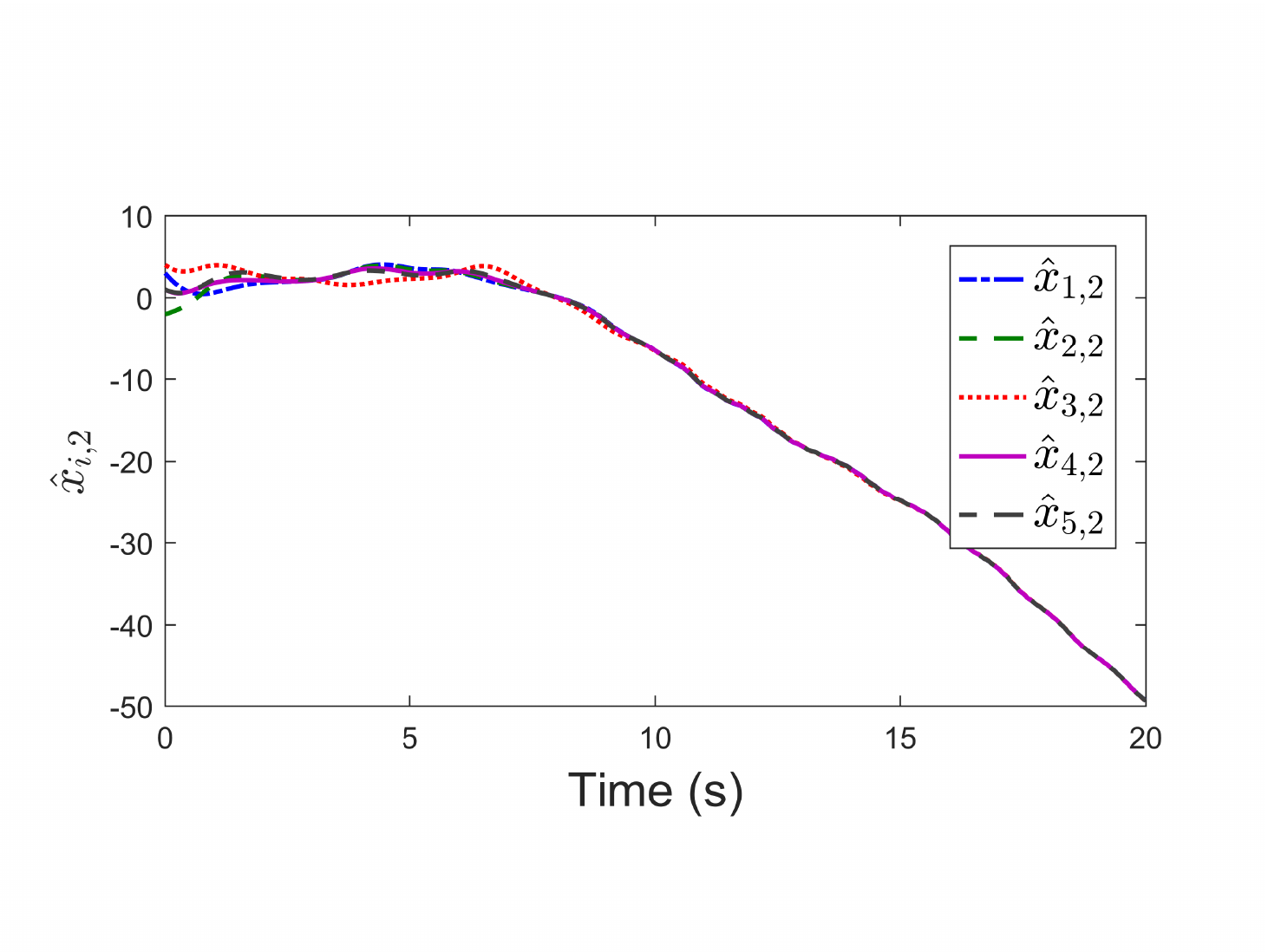}\label{Xi2-Nonli}} 
\subfigure[]{
\includegraphics[trim={5mm 16mm 12mm
 20mm},clip,width=0.325\linewidth]{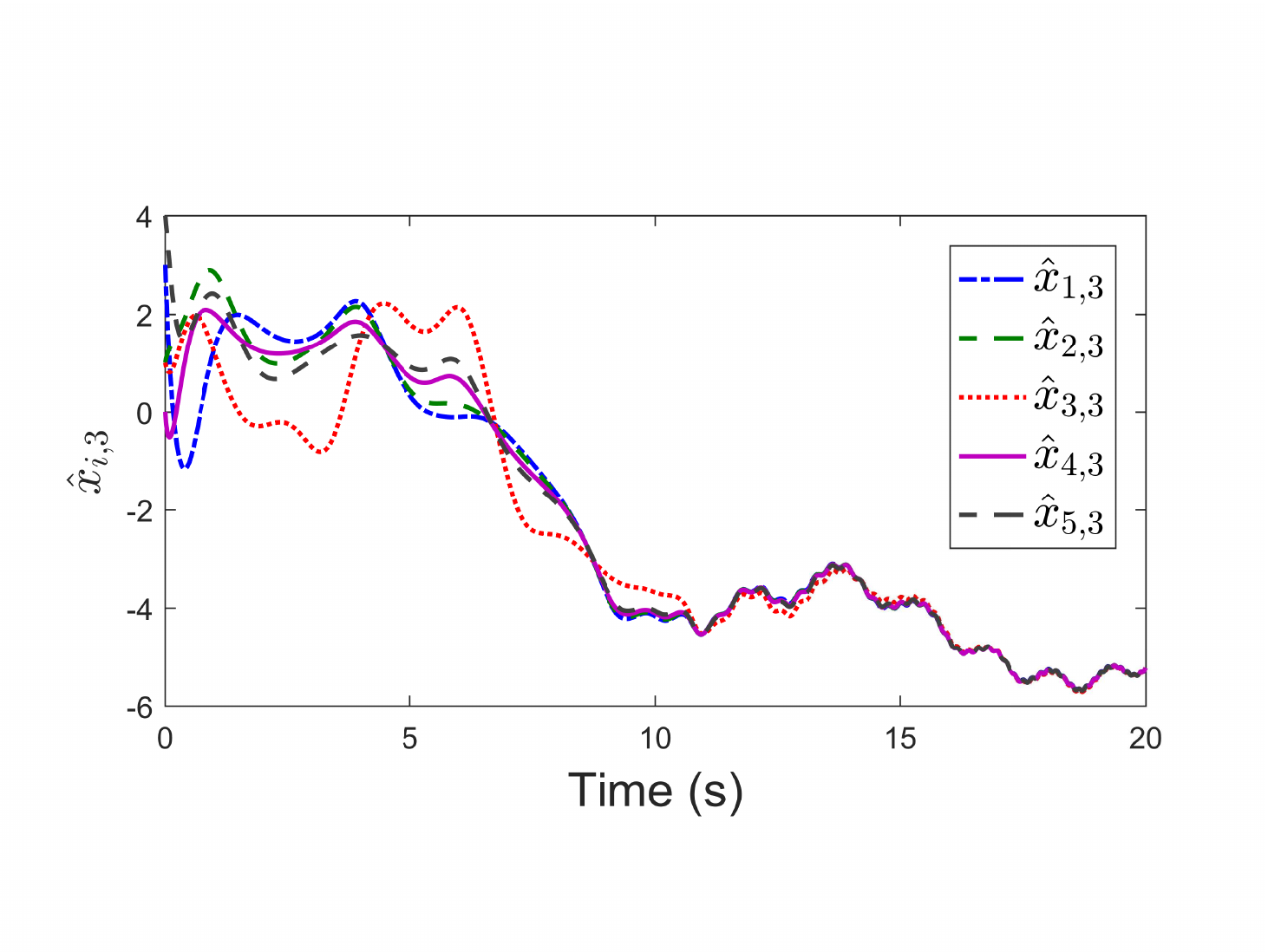}\label{Xi3-Nonli}}
\caption{Third-order nonlinear MAS profiles: (a) leader's and followers' state trajectory profiles of $x_{i,1}$ for $i=0,1,\ldots,N$; (b) leader's and followers' state trajectory profiles of $x_{i,2}$ for $i=0,1,\ldots,N$; (c) leader's and followers' state trajectory profiles of $x_{i,3}$ for $i=0,1,\ldots,N$; (d) leader's input profile and the estimation by each follower; (e)  leader's state trajectory profile of $x_{0,1}$ and the estimation by each follower;
(f) leader's state trajectory profile of $x_{0,2}$ and the estimation by each follower; (g) leader's state trajectory profile of $x_{0,3}$ and the estimation by each follower; (h) followers' estimation of their own state trajectories of $x_{i,2}$ for $i=1,2,\ldots,N$; (i) followers' estimation of their own state trajectories of $x_{i,3}$ for $i=1,2,\ldots,N$.}\label{Third-order-MAS-profiles-Nonli}
\end{figure*}

\section{NUMERICAL STUDY}\label{Simulation}

This section presents  a numerical simulation example to show the effectiveness of the proposed design. For the sake of space, we only illustrate the case of  nonlinear leader-follower tracking. Consider a third-order MAS including one leader and five followers. The agents interchange information based on a communication topology   shown in Fig.~\ref{MAS-topology-sim}. Here, node 0 is the leader, and nodes  1 to 5 are followers. The leader  transmits data to only follower 1, and the followers  maintain bidirectional communication with their neighbors.
The agents' dynamics is as described in~\eqref{High-Dynamics-Nonli}, for which   $ f_m(\overline{x_{i,m}})=\cos(\overline{x_{i,m}})^\top \mathbf{1} = \sum_{k=1}^m \cos(x_{i,k})$. 
The leader's maneuver input  is set to be
$
u_0 =\sin(0.2\pi t)
$. 
When implementing the proposed observer-based controller, we select $c_{0,1}=c_{0,2}=c_{0,3}=5$, $r_2=r_3=4$ and $k_1=k_2=k_3=3$. Such a gain set is verifiable to make   the convergence conditions  satisfied.
The simulation results are summarized in Fig.~\ref{Third-order-MAS-profiles-Nonli}.  Figs.~\ref{X01-Nonli}-\ref{X03-Nonli} illustrate followers' and the leader's state trajectories,   showing that the followers can manage to catch up with  and then keep tracking the leader, despite they differ in initial states. Fig.~\ref{U0i-Nonli} shows the estimation of the leader's input    by the followers. For each follower, the estimation can quickly converge to the actual values.
Meanwhile, the followers can also effectively  estimate the leader's states using the designed observer, with the estimation errors approaching zero as shown in Figs.~\ref{X0i1-Nonli}-\ref{X0i3-Nonli}. Figs.~\ref{Xi2-Nonli} and~\ref{Xi3-Nonli} further present the followers'   estimation of their own unmeasured states. These results validate that  the proposed design can  ensure   convergent tracking, despite the nonlinearity and limited information availability.

\section{Conclusion}\label{Conclusion}

We    studied   leader-follower tracking control for high-order MASs in this paper.   While this problem has recently attracted growing attention, the previous studies generally require  all the states of an agent to be measured, even though the measurement information can be practically limited by the availability of sensors. Here, we focused on the challenging but more realistic setting where  only the first state of an agent is measured.  
We designed novel distributed observers, by which a follower can reconstruct unknown or unmeasured quantities about itself and the leader, and then performed distributed observer-based controller synthesis. We conducted the design for both linear and nonlinear MASs and characterized the convergence properties. A simulation result demonstrated the effectiveness of our design.  Our future work will be directed towards extending the results to directed graphs and completely unknown nonlinearity.

\bibliography{reference}

\end{document}